\documentclass{svproc}
\usepackage{graphics}
\usepackage{url}

\def\mups{$M_{up}^{\star}$~}
\def\mup{$M_{up}$~}
\def\msun{$M_\odot$~}
\def\c12c12{$^{12}$C$+^{12}$C}

\begin{document}
\mainmatter              
\title{On the mass of supernova progenitors: the role of the $^{12}$C$+^{12}$C reaction}
\titlerunning{Mass of supernova progenitors}  
%
\author{Oscar Straniero\inst{1,2} \and Luciano Piersanti\inst{1,3}
Inma Dominguez\inst{4} \and Aurora Tumino\inst{5,6}}
\authorrunning{Oscar Straniero et al.} 
%
%
\institute{INAF, Osservatorio Astronomico d'Abruzzo, Teramo, Italy,\\
\email{oscar.straniero@inaf.it}
\and
INFN Laboratori Nazionali del Gran Sasso, Assergi, Italy
\and
INFN Perugia, Italy
\and
Departamento de Fisica Teorica y del Cosmos, Universidad de Granada, Granada, Spain
\and
Facolt\'{a} di Ingegneria e Architettura, Universit\'{a} degli Studi di Enna Kore, Enna, Italy
\and
INFN, Laboratori Nazionali del Sud, Catania, Italy
}

\maketitle              

\begin{abstract}
A precise knowledge of the masses of supernova progenitors is essential to answer 
various questions of modern astrophysics, 
such as those related to the dynamical and chemical evolution of Galaxies.
In this paper we revise the upper bound for the mass of the progenitors of 
CO white dwarfs (\mup) and the lower bound for the mass of the progenitors of
normal type II supernovae (\mups). In particular, 
we present new stellar models with mass between 7 and 10 \msun, discussing their final destiny 
and the impact of recent 
improvements in our understanding of the low energy rate of the \c12c12 reaction.   
 
\end{abstract}
\section{Introduction}
\mup is the minimum stellar mass that, after the core-helium burning, develops temperature and 
density conditions for the occurrence of a hydrostatic carbon burning. 
Stars whose mass is lower 
than this limit are the progenitors of C-O white dwarfs and, when they belong to a close binary system, 
may give rise to explosive phenomena, such as novae or type Ia supernovae. Stars whose mass is 
only slightly larger than \mup ignite C in a degenerate core and, in turn, experience a thermonuclear 
runaway. Their final destiny may be either a massive O-Ne white dwarf or an e-capture supernova. 
More massive objects ignite C in non-degenerate conditions and, after the Ne, O and Si burning,
 they produce a degenerate Fe core. These stars are the progenitors of various 
type of core-collapse supernovae among which the well known type IIp. 
In spite of their importance, a precise 
evaluation of \mup and \mups is still missing (see,e.g., \cite{doherty2015}, \cite{straniero2016}).
It relies on our knowledge of various input physics used in stellar 
modelling, such as the plasma neutrino rate, responsible of the cooling of the core, the equation of 
state of high density plasma, which affects the compressibility and the consequent heating of the 
core, and the \c12c12 reaction rate. In addition \mup and \mups depend on the C-O core mass, 
which is determined by the extension of the convective instabilities during the H and He-burning 
phases, such as convective overshoot, semiconvection or rotational induced mixing.
In this paper, 
we revise the theoretical predictions of \mup and \mups. New stellar models of stars whose initial 
mass ranges between 7 and 10 \msun and nearly solar composition, i.e., Z=0.02 and Y=0.27, are presented. 
Finally we provide a quantitative evaluation of 
the effects of the latest \c12c12 reaction rate as revised after the new experimental studies 
based on the Trojan Horse Method \cite{tumino2018}.  
\begin{figure}
\centering      
\resizebox{0.98\textwidth}{!}{\includegraphics{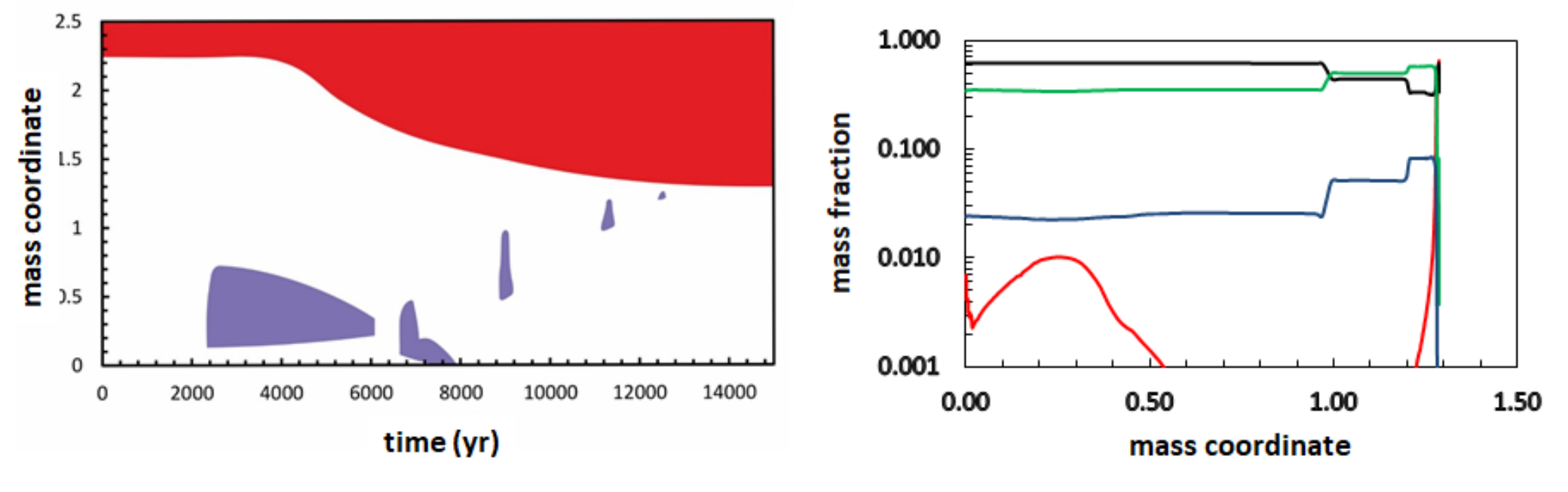}}
\caption{(Color online) Left: Kippenhahn diagram of the degenerate C burning in the 8.5 \msun stellar model. 
The red region corresponds to the convective envelope, while the violet regions are the convective 
C-burning episodes. Note that the t=0 point is arbitrary. 
Right: final composition: C (red), O (black), Ne (green), Mg (blue).}
\label{fig12}   
\end{figure}

\section{Carbon burning in degenerate C-O cores}
All the models here presented have been computed with the latest version of the FUNS code 
(see \cite{straniero2006} and \cite{piersanti2013}). 
Rotation is here neglected, but its effects 
will be discussed in a fothcoming paper. 
The Kippenhahn diagram of the C burning phase of the 8.5 \msun stellar model is shown in figure \ref{fig12} (left).
C ignites at about 0.15 \msun from the center, where the temperature attains its maximum value. 
More inside, the cooling induced by the emission of plasma-neutrinos prevents the C burning. Suddenly, 
a quite extended convective shell develops, powered by the thermonuclear runaway. This episode lasts 
for about 4000 yr. Then, for a short period of time the C burning dies down, meanwhile the contraction 
and the consequent heating restart, until a new off-center C ignition follows. This time, 
the released thermal energy diffuses inward, so that
the maximum temperature moves toward the center and a convective core develops. After about 500 yr, 
the convective core disappears.
Later on, the maximum temperature moves outside again, 
giving rise to 3 distinct convective  C-burning episodes.
 The final composition of the resulting O-Ne core 
is shown in figure \ref{fig12} (right). The main components are $^{16}$O and $^{20}$Ne, with a lower, but non-negligible,
 amount of  $^{24}$Mg. Note that in the innermost 0.5 \msun, the original carbon has not besn fully consumed.  
\begin{figure}
\centering      
\resizebox{0.78\textwidth}{!}{\includegraphics{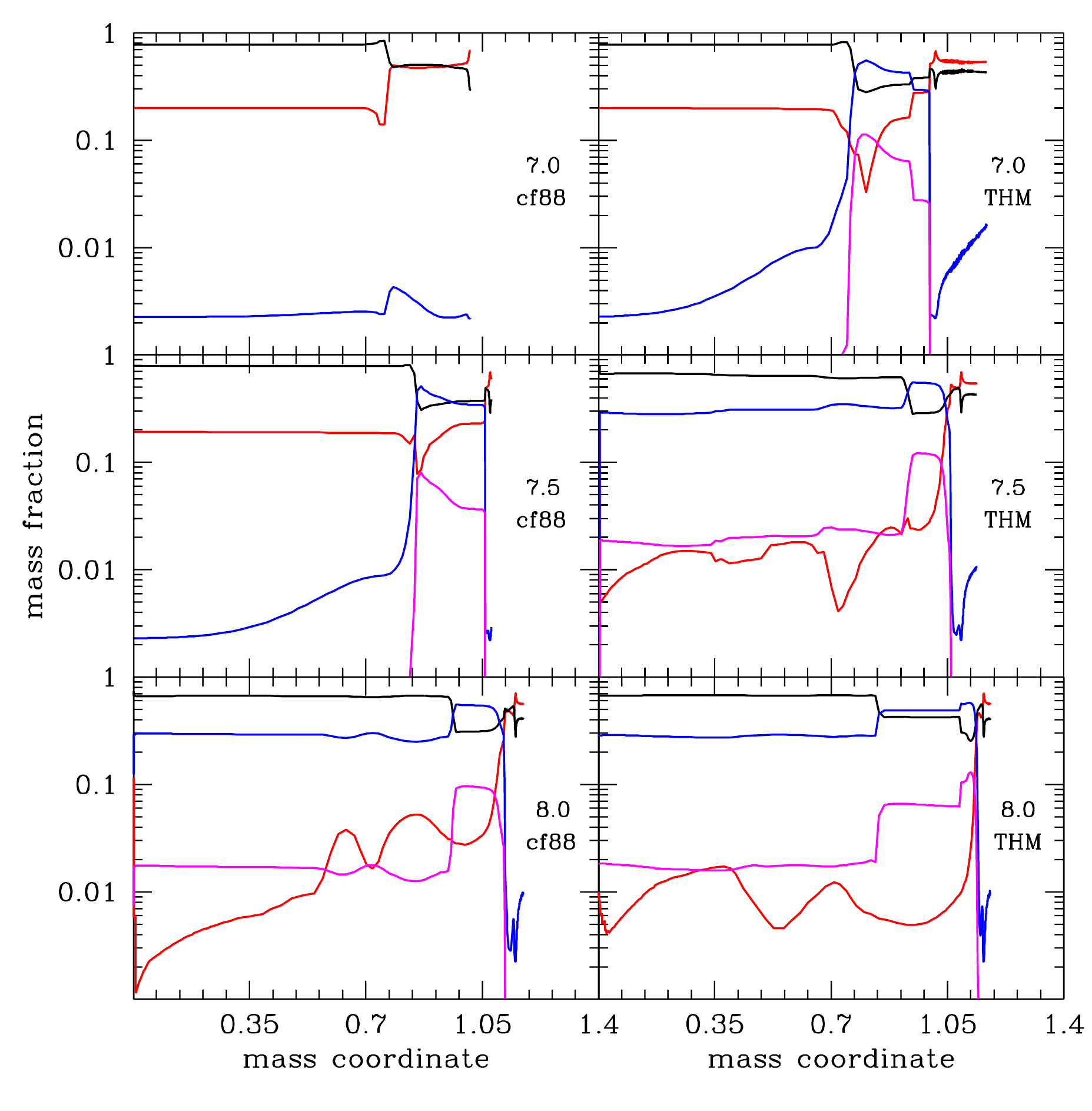}}
\caption{(Color online) Left: final composition within the core of the 7.0, 7.5 and 8.0 \msun stellar model computed by adopting
the CF88 rate for the \c12c12 reaction. Right: as in the left panels,
 but adopting the 
THM rate for the \c12c12 reaction. Colors represent different chemical species: 
C (red), O (black), Ne (blue), Mg (magenta).}
\label{fig3}   
\end{figure}

\section{\mup and \mups}
In order to determine the values of \mup and \mups, we have computed 2 sets of models, 
with masses ranging between 7.0 and 10.0 \msun (step 0.5 \msun), and different choices 
of the \c12c12 reaction rate, namely, the Caughlan and Fowler (CF88) \cite{cf88} and the 
Trojan Horse (THM) \cite{tumino2018} rates. 
Figure \ref{fig3} reports the final composition of the cores for the models 7.0, 7.5 and 8.0 \msun. 
With the CF88 rate (left panels), the lighter model never attains the conditions for the 
C ignition and proceeds its evolution entering the AGB phase. Then, 
the 7.5 and the 8.0 \msun undergo an incomplete C burning. 
In particular, in the 7.5 \msun
model, the C burning remains confined within a small shell located near the external border of the 
C-O core. Only for  $M\ge8.5$ \msun, after the usual off-center ignition, 
the C burning propagates inward down to the center, 
and a complete C burning takes place, resulting in the formation of an O-Ne core.
When the  THM rate is adopted (right panels in figure \ref{fig3}), we find similar final compositions 
but for models $\sim0.5$ \msun less massive.  
\begin{figure}
\centering      
\resizebox{0.78\textwidth}{!}{\includegraphics{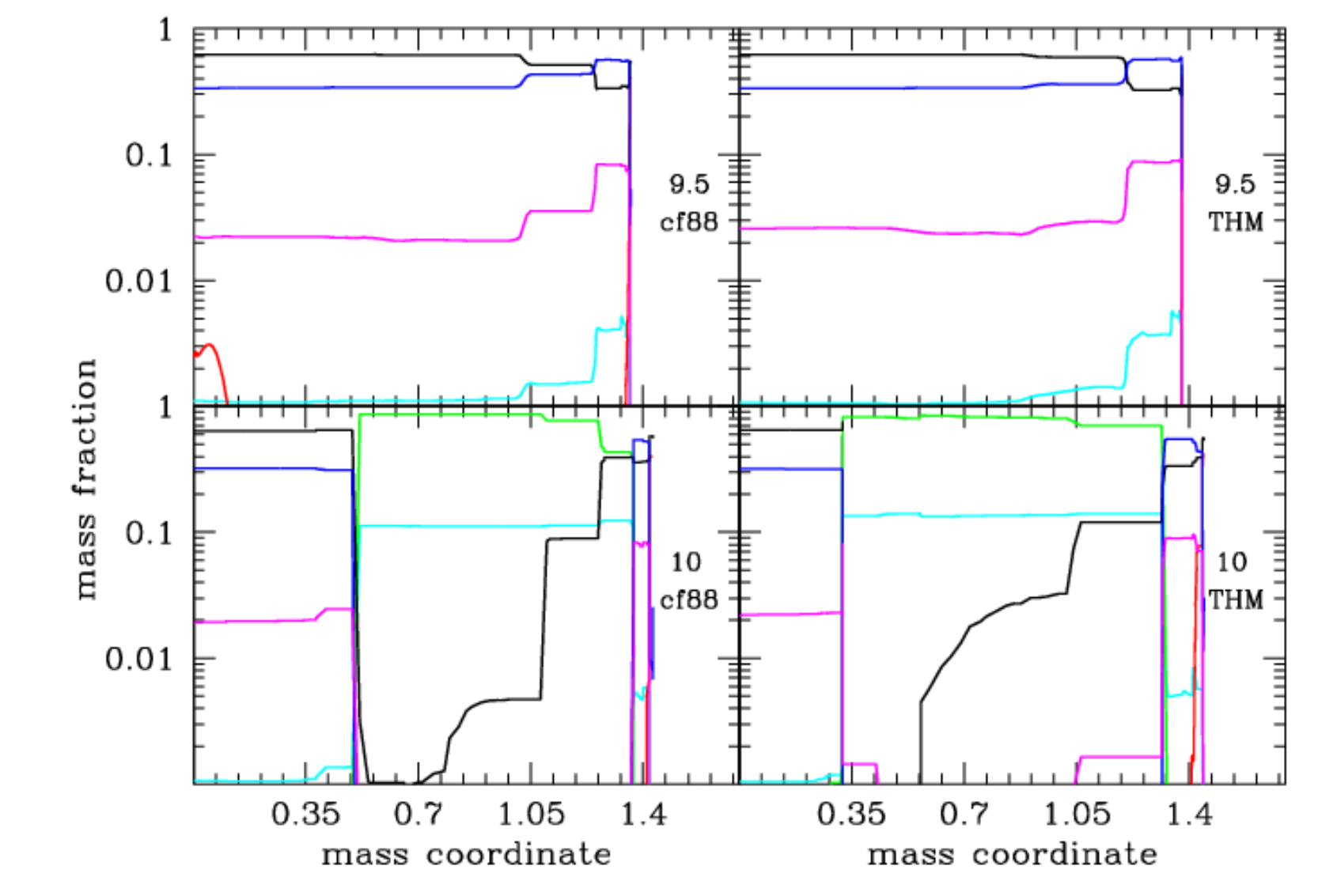}}
\caption{(Color online) Left panel: As in figure \ref{fig3}, but for M=9.5 and 10 \msun. Colors represents: 
C (red), O (black), Ne (blue), Mg (magenta), Si (green), S (cyan).}
\label{fig4}   
\end{figure}
Concerning \mups, figure \ref{fig4} shows the final compositions 
of the 9.5 and 10 \msun models. In this case we find little 
differences between the CF88 and the THM models. In both cases, the 
9.5 \msun model never attains the conditions for the Ne photo-dissociation. On the contrary
the 10 \msun models ignite C at the center. Then, 
they experience a complete C burning. Later on, the Ne photo-dissociation 
starts at about 1 \msun and the evolution proceeds through 
the more advanced burning phases (note that we stopped the computations when the O burning was 
still moving inward). 

%
%


\end{document}